# Human error in motorcycle crashes: a methodology based on in-depth data to identify the skills needed and support training interventions for safe riding

**Pedro Huertas-Leyva**[*], **Niccolò Baldanzini, Giovanni Savino, Marco Pierini**

*Dipartimento di Ingegneria Industriale, Università degli Studi di Firenze, Via di Santa Marta 3, 50139 Firenze, Italy*

[*]Corresponding Author:
pedro.huertasleyva@unifi.it

## Abstract

**Objective**: Human error is considered the primary factor contributing to crashes involving powered-two-wheelers (PTW), however not human-factors-based crash analysis methodology has been developed to support effectiveness of rider training interventions. Our aim is to define a methodology that uses in-depth data to identify the skills needed by riders in the highest risk crash configurations to reduce casualty rates.

**Methods**: The methodology is illustrated through a case study using in-depth data of a total of 803 powered-two-wheeler crashes cases. Seven types of high-risk crash configuration based on the pre-crash trajectories of the road-users involved were considered to investigate the human errors as crash contributors. Primary crash contributing factor, evasive manoeuvres performed, horizontal roadway alignment and speed-related factors were identified, along with the most frequent crash configurations and those with the greatest risk of severe injury.

**Results**: Straight Crossing Path/Lateral Direction was the most frequent crash configuration and Turn Across Path/ Opposing Direction that with the greatest risk of serious injury were identified. Multi-vehicle crashes cannot be considered as a homogenous category of crashes to which the same human failure is attributed, as different interactions between motorcyclists and other road users are associated with both different types of human error and different rider reactions. Human error in multiple-vehicle crashes related to crossing paths configurations were different from errors related to rear-end or head-on crashes. Multi-vehicle head-on crashes and single-vehicle collisions frequently occur along curves. The involved collision avoidance manoeuvres of the riders differed significantly among the highest risk crash configurations. The most relevant lack of skills are identified and linked to their most representative context. In most cases a combination of different skills was required simultaneously to avoid the crash.

**Conclusions**: The results contribute to understand the motorcyclists' responses in high-risk crash scenarios. The findings underline the need to group accident cases, beyond the usual single-vehicle versus multi-vehicle collision approach. The different interactions with other road users should be considered in order to identify the competencies of the motorcyclists needed to reduce the risk of an accident. Our methodology can be applied to increase the motorcyclists' safety by supporting preventive actions based on riders' training and eventually ADAS design.

# INTRODUCTION

Each year, about 1.35 million people die in the world as a result of road traffic crashes, and users of motorcycles and mopeds, together referred to as powered-two-wheelers (PTWs), represent nearly a quarter of the fatalities. A fundamental issue to be addressed in the analysis of the PTW crash causes is the human factor contribution, considered the primary crash contributor in in-depth studies conducted worldwide (ACEM 2009; Hurt et al. 1981).

To reduce the number of crashes caused by human failure, researchers proposed science-based countermeasures aiming at improving riding competencies and behaviour through effective training and sensitization campaigns (Haworth and Rowden 2010). However, there is little evidence about the effectiveness of any training program in reducing crash risk (Ivers et al. 2016). The fact that some studies have reported the ineffectiveness of certain rider training programs does not necessarily indicate the inefficacy of training to increase safety, but rather the need for more effective training design.

Given the overall goal of minimizing PTW crash risks, chances of successful training would derive by addressing the key riding skills needed in those crash types with the highest risk. Some authors have proposed that these skills can be identified with in-depth crash investigations through a comprehensive analysis of the crash contributing human errors (Clarke et al. 2007; Salmon et al. 2010).

Previous in-depth studies of PTW crashes have mainly evaluated the effect of crash contributing factors on both crash incidence (Vlahogianni et al. 2012) and injury severity (Savolainen and Mannering 2007). Previous research on human errors in PTW crashes is insufficient, primarily due to two limitations. First, most of the available crash-data lack detailed and consistent information on human responses and related errors (Van Elslande, 2002, Salmon et al.2010), which limits the definition of specific human-factors based countermeasures. Secondly, in the few cases where data contained detailed human responses information, the analyses used the crash cases as one homogeneous group or, at best, grouped the cases into single-vehicle (SV) or multi-vehicle (MV) crashes (Allen et al. 2017). Such analyses provide a broad view of human error in crashes, but overlook the effect of certain crash characteristics on those errors (Mannering and Bhat 2014), such as the interaction between the road users involved. Crash countermeasures based on this type of analysis assume that the identified human errors affect all crash scenarios similarly. Consequently these analysis are more sensitive to biases in the data collected caused by factors such as the road type or the geographical location (e.g. crashes collected in urban roads will occur more often both at intersections and at lower speeds than those collected in rural areas).

To overcome the above limitations, we propose a crash analysis methodology that identifies the essential competencies needed to reduce the incidence of each of the most frequent crash types. This methodology uses detailed data on human response and considers the type of crash configuration as an explanatory variable of the interaction between involved road-users in the crash. We address the following two hypothesis: (1) PTW crashes with different types of interaction between the involved road-users are connected to different types of human errors; (2) Riders need to combine different type of skills contemporary to avoid the most common type of crashes.

# METHODS

## Data

PTW crash data collected in France, Germany, Netherlands, Spain and Italy under the MAIDS project (ACEM 2009) were analysed. The sample contained 921 crashes cases with injured PTW users (including 100 fatal crashes) between 1999 and 2001, with 1721 variables reported. For our analysis we excluded crashes where the primary contributing factor was rider impairment or mechanical problems (i.e. rider or PTW were not in a riding condition) and PTW mofa-type crashes, reducing the sample to 803 cases (see Appendix Figure A1). Additional references can be found in the bibliography in the Appendix.

## Crash Types Selection

We selected the 16 configurations of the 25 included in MAIDS (see Appendix Table A1), where either an evasive manoeuvre was required or a simple collision occurred without involving other road users (665 cases). To facilitate the identification of human error by crash configuration and increase the statistical power, we merged those configurations that shared similarities in pre-crash trajectories of the vehicles. The resulting seven configurations (Figure 1) correspond to: four with multi-vehicle (MV) crossing path crash at intersections; two with MV collisions not related to intersections; and one single vehicle (SV) crash with the PTW falling or running off the roadway with no other vehicle (OV) involved. The remaining configurations were categorized as *'Other'* (138 cases).

## Crash Variables to Determine Lack of Competencies

First, frequency distribution of each crash configuration was analysed as whole and segmented by *PTW category* (*Moped* or L1 for PTWs with speed ≤45km/h and engine ≤50cm³; *Motorcycle* or L3 for higher speed and engine) and by *Injury severity* (*severe* if MAIS3+ and *non severe* otherwise).

**Figure 1. Definition of the seven merged crash configurations selected and the variables used for crash characterization.**

Secondly, the type of human errors involved and the attempted avoidance responses were identified by analysing: the categorical variables (*i*) *primary crash contributing factor, (ii) horizontal roadway alignment, (iii) evasive manoeuvres;* and the quantitative variables *(iv)* Time from Precipitating Event to Impact (*TPEI*); and (v) *posted speed limits* and estimated *impact speed* (Figure 1).

- *Primary crash contributing factor,* identified as generator of the crash event, included *environment factors* (road surface, adverse weather and view obstruction) and *human failures*. The classification of human failures, conducted in agreement with the stages of the information processing (Stanton and Salmon 2009; Wierwille et al. 2002), were:
    i. late *detection* error (listed as *perception* in MAIDS), including cognitive errors (do not look at the area of the hazard) and perception errors (looked-but-failed-to-see);
    ii. *comprehension (*poor diagnostic of the scene*);*
    iii. *decision* (manoeuvre or response selection at a wrong time or place);
    iv. *execution* (listed as *reaction* in MAIDS), relate mainly with skill-based errors – e.g. overcompensation, inadequate control or freezing.
- *Evasive manoeuvers* were categorized by *braking* (activation of rear, front or both brakes), *swerving*, *no evasive action* and *other* (including flashing headlamp high beams, accelerating or honking). Error in the *selection* and *execution* of the attempted manoeuvre was also analysed.

## Statistical Analysis

Descriptive statistics by crash configuration included frequency distributions of the categorical variables and mean and quartile 1 and 3 (Q1 and Q3) for *posted speed limits*, estimated *speed impact* and *TPEI*. When impact speed exceeded the posted speed limit by 20%, we considered *speeding* as a crash contributing factor.

Odds ratios (OR) with their 95% confidence intervals were calculated to determine the association between each type of crash configuration and the variables *injury severity*, *PTW category* and *evasive manoeuvre*. For OR analysis, the original categorical variables were recoded as sets of dichotomous variables (category present or absent). Each analysis computed the risk of the analysed crash scenario versus the pooled remaining scenarios. Statistical significance was considered when 1.0 was out the 95%CI.

## RESULTS

### Frequency Distribution

Table 1 shows the frequency distributions for the seven crash configurations considered and for those crashes categorized as *Other*. Overall, the most frequent PTW crash configurations were SCP/LD (16.9%) and TAP/SD (16.7%). The four configurations that gather the MV crossing path crashes at intersection (SCP/LD, TIP/LD, TAP/OD and TAP/SD) represented the 57.9% of all the crashes. The two configurations representing MV collision not related with intersections (RE/SD and HS/OD) were the least frequent (less than 8% each) while single vehicle crashes without OV involvement (SV) represented 11.1% of the PTW crashes.

For the motorcycle PTW type subset, the most frequent crash configuration was SV with 16.1% of the cases. Motorcycles were overrepresented in SV configuration compared to the rest of MV crash configurations (87.6% vs. 56.7%). SV represents a risk for motorcycles 5.4 higher than for mopeds (OR=5.4, 95%CI [2.8, 10.3]). Severe injuries were significantly more frequent in TAP/OD crashes (32.6%) than in all other configurations put together (21.3%). Crashes in TAP/OD represented a risk of severe injury 1.8 higher than in all other configurations put together (OR = 1.8, 95%CI [1.1, 2.9]).

**Table 1. Frequency of crashes by crash configuration. (N=803 cases)**

| Crash Configuration | | Total (N=803) | Severe Injury (N=182) | No Severe Injury (N=618) | L3 Motorcycle (N=483) | L1 Moped (N=320) |
|---|---|---|---|---|---|---|
| 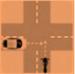 | SCP/LD* | **16.9%** | 14.3% | 17.8% | 13.9% | **21.6%** |
| 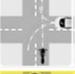 | TIP/LD | 12.5% | 14.8% | 11.7% | 10.8% | 15.0% |
| 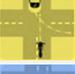 | TAP/OD† | 11.8% | **17.0%** | 10.4% | 13.0% | 10.0% |
| 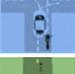 | TAP/SD* | **16.7%** | 12.1% | **18.1%** | 14.9% | 19.4% |
| 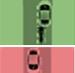 | RE/SD | 6.5% | 5.5% | 6.8% | 7.0% | 5.6% |
| 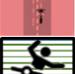 | HS/OD | 7.3% | 7.7% | 7.0% | 6.4% | 8.8% |
| 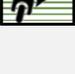 | SV‡ | 11.1% | 11.5% | 11.0% | **16.1%** | 3.4% |
| | OTHER | 17.2% | 17.0% | 17.3% | 17.8% | 16.3% |
| | TOTAL | 100% | 100% | 100% | 100% | 100% |

*most frequent overall;  † most frequent for severe injury; ‡ most frequent for L3; + most frequent for no severe injury

## Primary Contributing Factor

Figure 2 shows that distribution of primary crash contributing factors for the whole dataset (*Total*) differs from most of the scenarios and only presents similarities with the two more frequent configurations (SCP/LD and TAP/SD) (more details about *other* factor in Appendix Table A2).

| | | Total | SCP/LD | TIP/LD | TAP/OD | TAP/SD | RE/SD | HS/OD | SV |
|---|---|---|---|---|---|---|---|---|---|
| **OV** | detection | 40.8% | **47.8%** | 56.0% | **67.4%** | 50.0% | 21.2% | 13.6% | 0.0% |
| | decision | 12.2% | 12.5% | 16.0% | 15.8% | 15.7% | 9.6% | 11.9% | 0.0% |
| | exec./compreh. | 1.7% | 1.5% | 0.0% | 2.1% | 3.0% | 1.9% | 3.4% | |
| **MC** | detection | 9.3% | 9.6% | 2.0% | 3.2% | **11.9%** | 28.8% | 8.5% | 9.0% |
| | comprehension | 2.4% | 3.7% | 3.0% | 2.1% | 3.0% | 3.8% | 0.0% | 0.0% |
| | decision | 13.1% | **11.8%** | 9.0% | 3.2% | **14.2%** | 11.5% | 25.4% | 21.3% |
| | execution | 4.8% | 1.5% | 0.0% | 1.1% | 0.0% | 5.8% | 13.6% | 20.2% |
| | unknown type | 2.4% | 0.0% | 0.0% | 0.0% | 0.0% | 0.0% | 1.7% | 16.9% |
| | view obstruction | 4.4% | 4.4% | 12.0% | 4.2% | 0.7% | 1.9% | 6.8% | 1.1% |
| | other | 9.0% | 7.4% | 2.0% | 1.1% | 1.5% | 15.4% | 15.3% | **31.5%** |

**Figure 2. Relation between *Primary crash contributing factor* and *Configuration* (N= 663; two cases missed for this variable). MC: PTW rider; OV: Other vehicle driver.**

*Multi-vehicle collision at intersection: Detection failure by other vehicle driver*

In the configurations corresponding to MV crossing path crashes at intersection (SCP/LD, TIP/LD, TAP/OD and TAP/SD), the primary contributing factor is mostly related to OV driver failure, including *detection* as the most frequent failure. This trend is particularly noticeable in TAP/OD where OV driver failure contributed to 85.3% of crashes, of which 67.4% corresponded to *detection* failure. PTW rider failure still had a relevant prevalence in two crash configurations at intersections SCP/LD and TAP/SD, where failures in *detection* and *decision* together make up more than 20% of the cases. *View obstruction* was a frequent contributing factor in crash configuration TIP/LD (12% of cases).

*Multi-vehicle collision without crossing paths and Single vehicle crashes: PTW rider failure*

In the two MV crash configurations not related to intersections (RE/SD, HS/OD) and in single vehicle crashes (SV) the primary crash contributing factor was *PTW rider failure* (50.0%, 49.2% and 67.4% of cases respectively). The most common failures were: in RE/SD *detection* (28.8%); in HS/OD *decision* (25.4%) and *execution* (13.6%); in SV *decision* (21.3%) and *execution* (20.2%) (e.g. wrong trajectory bending a curve). In SV, other external factors such as *roadway maintenance defects* (6%) or *adverse weather* (8%) included in the 31.5% of '*other*' category are also important.

## Horizontal Roadway Alignment

Crashes occurred with PTW circulating straight in over 74% of the cases in all crash configurations except for SV (24%) and HS/OD (34%) which occurred more frequently in curves. Left-hand curves were more prevalent in SV (left-hand 45% and right-hand 30%) and right-hand curves in HS/OD (left-hand 12% and right-hand 46%) (see Figure A2 in Appendix).

## Evasive Manoeuvres

Overall, *braking* is the most attempted evasive response (47%) followed by *no evasive action* (35%) and *swerving* (13%) (Figure 3).

*Braking* was overrepresented in TIP/LD (O.R.: 2.0; 95%CI [1.3, 3.0]), TAP/OD (O.R.: 1.5; 95%CI [1.0, 2.2]) and RE/SD (O.R.: 1.7; 95%CI [1.0, 3.0]). *Swerve* was overrepresented in TAP/SD (O.R.: 1.8; 95%CI [1.1, 3.0]), *no evasive action* was overrepresented in SCP/LD (O.R.: 1.7; 95%CI [1.2, 2.4]) and *other* action was over-represented in HS/OD (O.R.: 3.1; 95%CI [1.2, 7.8]).

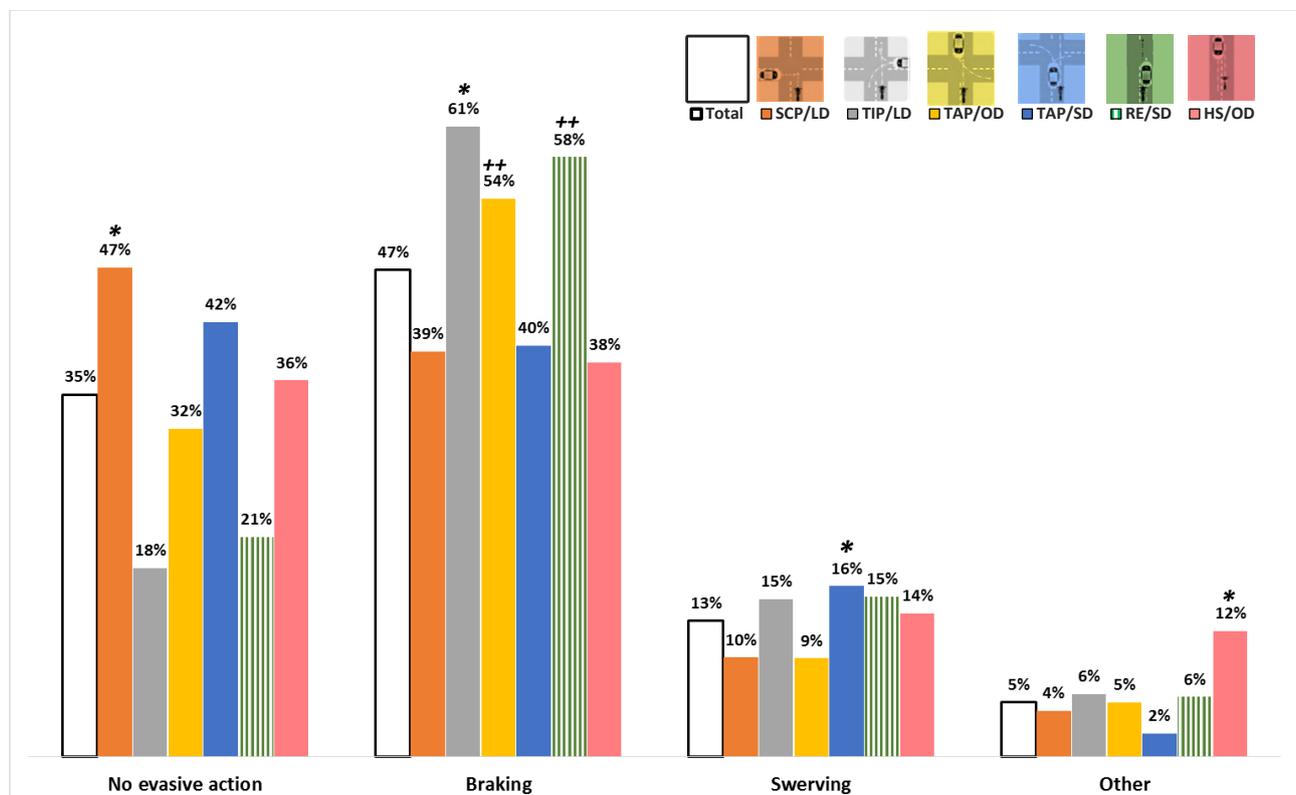

**Figure 3. Relation between *Collision Avoidance Manoeuvre* and Crash Configuration (N=576).**
  *\* 1.0 out of 95%CI; \+\+ 1.0 as the lower limit of 95%CI*

According to the in-depth data, excluding the *no evasive action* cases, the *selection* of the evasive manoeuver was correct in 81.6% of the 374 cases (76.4% was the lowest rate of correct manoeuvres for SCP/LD and 87.8% the highest for RE/SD. Therefore, we can assume that the connections found between evasive manoeuvres and crash configurations are not biased by *selection* failure. Analysing the cases with correct selection (N=305), *execution* failures were found in braking and swerving in 43% and 38% of the cases respectively, without significant effect of the crash configuration, except HS/OD where the risk of swerving wrongly was much higher (O.R.:14.9; 95%CI [1.7, 125.0]). Summarizing, when an evasive manoeuvre was attempted, incorrect selection was not a crash contributing factor; in contrast, poor braking or swerving often contributed to not avoiding the crash (see more information in Appendix Tables A3 A4).

## Speed and Time from Precipitating Event to Impact (TPEI)

*Posted speed limits* for the collision cases was 50km/h in most of the cases for all the configurations except for SV, where 100km/h was also predominant (Table 2). Posted limits are related with road type, in fact the data collected corresponds mostly to urban road where limits of 50km/h are predominant. The mean of estimated *impact speed* was 44.9km/h overall, with means ranging from 35 to 50 km/h in all crash configurations except for SV (63.8km/h). Overall, in 14.6% of the crashes the *impact speed* was 20% in excess of *posted speed limit*, with SV configuration having the highest prevalence of this type of speeding (21.2%), followed by TAP/OD (17.9%). Concerning *TPEI*, configurations SV, HS/OD and TAP/SD were those with the shortest time to perform the necessary processes for collision avoidance. (More data in Appendix Figure A3).

**Table 2. Mean [Q1-Q3] for *posted speed* limits and estimated *speed impact* (km/h) and for *TPEI* (Time estimated from Precipitating Event to Impact) (s). In bold the highest speed and lowest TPEI.**

| Crash Conf. | Posted Speed | N | Speed Impact | N | TPEI | N |
|---|---|---|---|---|---|---|
| SCP/LD | 50.5 [50–50] | 134 | 38.7 [25-49] | 136 | 1.9 [1.2- 2.4] | 136 |
| TIP/LD | 48.6 [40-50] | 99 | 35.6 [23-44] | 100 | 2.0 [1.4- 2.4] | 96 |
| TAP/OD | 55.5 [50-70] | 95 | **49.6** [30-70] | 95 | 2.0 [1.2- 2.1] | 94 |
| TAP/SD | 54.0 [50-50] | 134 | 43.4 [27-53] | 134 | **1.7** [1.0-2.4] | 133 |
| RE/SD | 59.9 [50-70] | 51 | 40.2 [26-51] | 52 | 2.2 [1.4-2.7] | 52 |
| HS/OD | 59.8 [40-90] | 59 | **47.1** [28-64] | 59 | **1.7** [1.1-2.1] | 58 |
| SV | **71.7** [50-100] | 86 | **63.8** [34-89] | 88 | **1.5** [0.6-2.3] | 83 |
| **Total** | 56.0 [50-50] | 658[†] | 44.9 [27-54] | 664[†] | 1.9 [1.1-2.3] | 652[†] |

[†] Sample less than n=665 due to some cases not having the required information

## DISCUSSION

This paper has examined in-depth data of 803 PTW injury crashes, grouped in crash configurations defined by the trajectories and interactions of the road users involved. The aim of this study was to define a methodology to identify both the human failures that contributed to crash and, consequently, the skills required in the highest risk crash configurations in the perspective of training interventions to enhance motorcyclist safety. We present a methodology based on: the identification of the most representative crash configuration and the comprehensive analysis of the rider behaviour using in-depth data. This work revealed that (a) different contexts are associated with different human failures and (b) a single training component is insufficient to reduce the occurrence of accidents in most high-risk traffic conflicts.

The first main finding showed that crash scenarios with different interactions between road-users involved are associated with different kind of human errors and riders' reactions. Multi-vehicle (MV) crashes cannot be considered as a homogenous type of crash with same human failures attributed in each MV crash configuration. The results warn about the likely inaccuracy of a human error analysis based on overall crash data without considering the different characteristics of the crash configurations. Our research shows the importance of categorising crash cases considering the different contexts and road users interactions to fully understand the human failure that contributed to collision.

The study found two clusters of configurations concerning the primary crash contributing factors. First cluster corresponds to the four MV collision configurations at intersections (SCP/LD, TIP/LD, TAP/OD and TAP/SD), and presented factors related to failure of other vehicle (OV) drivers as the most frequent primary contributing factor. *Detection* failure by OV drivers was the most common primary factors contributing to crossing path crashes. Different causes have been linked to other driver failure in PTW detection: PTW low conspicuity (Rößger et al. 2015); spatial frequency of the motorcycle; light density of traffic (Allen et al. 2017); and perception failure of the gap with the PTW of other driver before getting in the intersections (Horswill et al. 2005). According to the human failure model defined by Van Elslande (2002) in crashes with OV driver detection failures, the prognosis error should be also included as a PTW rider failure due to misjudgement or wrong anticipation (e.g. right-of-way false assumption). Despite the similitudes among the MV crash configurations of this cluster, we also found that these four configurations differed from each other either in the rider responses to avoid collisions or in the average *impact speed*, which suggests analysing these types of crashes independently.

Second cluster, that corresponds to the two MV configurations which are not crossing path crashes (PTW impacting rear of other vehicle in same direction: RE/SD; Head-on/Swipe with other vehicle in Opposite Direction: HS/OD) and the single-vehicle crash configuration without involvement of other road user (SV), presented factors related to rider failure as the most frequent primary contributing factor. *Detection* failure by rider due to late reaction (cognitive and/or perception failures) was the main crash contributor factor in the MV crash configuration RE/SD, and *decision* and *execution* failures by rider in SV and HS/OD crash configuration. *Decision* failures are connected with manoeuvres like overtaking due to a misjudgement or violations and speeding. Excessive speed was estimated in SV configuration in 21% of the cases. Time from precipitating event to impact (TPEI), a parameter indirectly related to speed, presented the lowest values in HS/OD, which in turn reduces the likely of a successful evasive manoeuvre. *Execution* failures in both SV and HS/OD may be related to a poor performance taking curves. Crashes in HS/OD were frequently (46% of cases) on right-hand curves that runs the PTW into oncoming traffic (traffic drove on the right), probably due to inappropriate lane position (Crundall et al. 2012). Crashes in SV were frequently on left-hand curves (45% of cases) that run the PTW off the roadway.

The second main finding showed that motorcycle riding is a complex task and that in most cases the combination of different skills was required to avoid the crash (Table-3). Training programs focusing on one single skill component, either cognitive or control skills, are likely to achieve a limited effect in reducing rider casualties. Rather, training should address multiple skills considering the different high-risk configurations, identifying the set of skills potentially able to reduce the likelihood of being caught in a particular crash scenario (see Figure A4 in Appendix).

Braking was the most commonly used evasive manoeuvre in most of the configurations, and in many cases the execution was considered poor, suggesting a need to improve avoidance response skills while avoiding loss of control (Huertas-Leyva et al. 2020). However, acquiring these control skills may be not enough to avoid crash if time to collision is too short. Considering that the average for TPEI is 2s (M=1.9s for our dataset) and for travelling speed is 50km/h, the average deceleration required to avoid the collision (in an emergency braking manoeuvre) should be higher than 6.94m/s$^2$. According to experiments from literature on emergency braking in a controlled environment, this deceleration is only achieved by experts (Huertas-Leyva et al. 2019). The significant proportion of cases found where riders did nothing to avoid collision may be related to different factors such as: how unexpected the hazard stimulus is (e.g. wrong anticipation); how many solutions the rider has under consideration; or how fast the perception-action process is under constrained time. Less experienced riders or with less automatism hazard-perception/evasive-action may get into panic mode, instinctively reacting with either freeze or fear responses (Collet et al. 2005). The findings support training anticipatory skills and tasks coupling perception and action to reduce cases with *no action* avoidance manoeuvres and to increase the safety margin. The perception-action task was defined as critical in previous naturalistic study with cars, which found that the main difference between crashes and near-crash cases was not in the main conflict contributor factor but in how quick was the evasive manoeuvre response after the precipitating event onset (Guo et al. 2010). Results also support training riders to select both proper lane position and speed in sharp curves. Finally, addressing risky attitudes, identifying the consequences of speeding and inappropriate overtaking decisions may also comprise a reduction of motorcyclists' overconfidence and a change in attitudes and behaviour (Mannering and Grodsky 1995). Concurrently, the importance of driver education in learning to perceive PTWs at intersections should be stressed.

**Table-3. Summary of crash configurations features based on human error analysis**

| Config. | Prim Cont | Horizontal Road Alig. | No Evasive >25% cases | Evasive Over-represented | Speed Impact (mean) | Other |
|---|---|---|---|---|---|---|
| **SCP/LD** 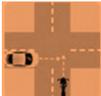 | OV Failure: *Detection* | Straight | No action (47%) | No action (wrong anticipation) | 38.7Km/h Urban road | Overall PTW: Most frequent (16.9%) |
| **TIP/LD** 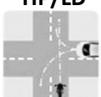 | OV Failure: *Detection* | Straight | | Braking | 35.6 km/h Urban road | View obstruction (12%) |
| **TAP/OD** 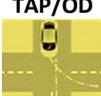 | OV Failure: *Detection* | Straight | No action (32%) | Braking | 49.6 Km/h (18% 'speeding') | 17.0% of Severe injury crashes highest risk * |
| **TAP/SD** 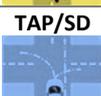 | OV Failure: *Detection* | Straight | No action (42%) | Swerve (over) | 43.4 Km/h Urban road | Overall PTW: high frequency (16.7%) Mean(TPEI) ≤1.7s |
| **RE/SD** 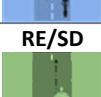 | MC Failure: *Detection* | Straight | | Braking | 40.2 km/h | TPEI: Highest Q3 (2.7s) |
| **HS/OD** 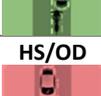 | MC Failure: *Decision* (overtaking) *Execution* (curve trajectory) | Curve (Righ-hand) | No action (36%) | Other (over) Poor Swerve* | 47.1 km/h | Mean(TPEI) ≤1.7s |
| **SV** 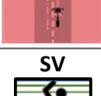 | MC Failure: *Decision* (speed) *Execution* (curve trajectory) | Curve | | | 63.8 Km/h (21% speeding) | L3: Most frequent (16.1%) Mostly L3 category* |

*significant level*

This study has a number of limitations. The data used may not be up-to-date and not accurately represent the current context (higher ratio of PTWs with advanced safety systems, infrastructure improvements or changes in riders' profile). Data were mainly collected in urban areas with a high proportion of mopeds, so the crash prevalence of the analysed configurations may have been affected by the influence of exposure. A fully developed approach to rider failures should include elements such as motivation, expectations and physical condition (Van Elslande, 2002), which can hardly be collected. MAIDS data, despite limitations, still represent one of the most comprehensive and complete in-depth databases on PTW crashes available in Europe today, including human failure analysis and rider response in each crash case, and with sufficient cases to support statistics analysis of multiple crash configurations. The proposed methodology can be applied to in-depth investigations with new data. Additionally, the approach presented, which uses information from road-user interactions to categorise crash configurations, allows comparing rider errors linked to each crash configuration with other databases that follow our same approach, avoiding the bias of an evaluation of human failure from a global approach (more sensitive to the overrepresentation of a particular crash scenario). Our methodology can be applied with in-depth studies of different geography to assess whether different populations need improving same or different competencies for safety. For a reliable comparison between studies in the future, it is critical to harmonise data collection criteria regarding the categories and terminology of both contributing factors and accident typologies used to classify crashes.

The results from the study give some keys in order to associate training skills required with representative test scenarios of the variability of real-world crashes. The importance of identifying the lack of competencies in specific context lies in their application to new training programs and education interventions for improving motorcycle safety. The findings can provide a valuable tool for motorcycle training, where instructors will know the most realistic scenario linked to a set of competencies identified as necessary to avoid rider failures in traffic conflicts. A content where competencies are learnt in a context of specific real-world crash scenarios may enhance rider motivations and consequently could improve the skills acquisition. Accordingly, instructors will have the possibility to adapt the specific skills training exercises to their most appropriate test scenario with the means available to them (videos, multimedia, simulator or off-road training). Additionally, the methodology presented, following previous studies on car drivers (Nilsson et al. 2018; Stanton and Salmon 2009),  may be applied to the definition of ADAS or ITS for PTWs aimed at strengthening riders' competencies and reducing frequent errors.

## ACKNOWLEDGEMENT

This work was funded by the 7th Framework Program of the European Commission within the Marie Curie Research Training Network MOTORIST (MOTOrcycle Rider Integrated SafeTy, grant agreement n. 608092).

We thank ACEM (European Motorcycle Manufacturers Association) for their availability by giving us access to the MAIDS database at their facilities in the framework of the MOTORIST project.

# APPENDIX

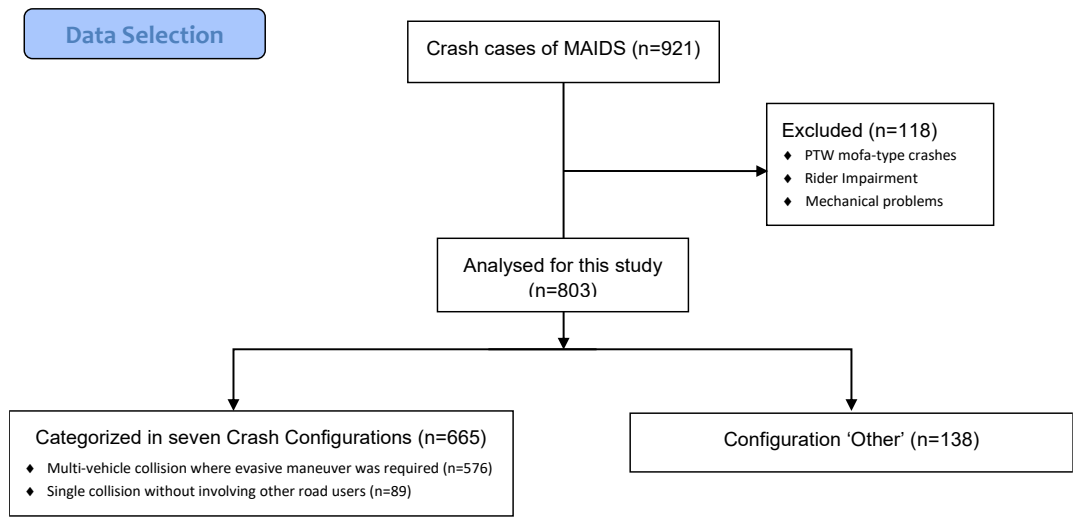

**Figure A-1 Process of data selection of the study**

**Table A- 1. Crash configuration grouping configurations from MAIDS**

| Crash Configuration | | 25 Crash group configurations defined by MAIDS |
|---|---|---|
| 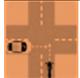 | SCP/LD* | PTW into OV impact at intersection; paths perpendicular |
| | | OV into PTW impact at intersection; paths perpendicular |
| 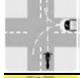 | TIP/LD | OV turning left in front of PTW, PTW perpendicular to OV path |
| | | OV turning right in front of PTW, PTW perpendicular to OV path |
| 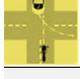 | TAP/OD† | PTW & OV in opp. dir., OV turns in front of PTW, PTW impacting |
| | | PTW & OV in opp. dir., OV turns in front of PTW, OV impacting |
| 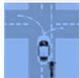 | TAP/SD* | PTW overtaking OV while OV turning left |
| | | OV making U-turn or Y-turn ahead of PTW |
| | | sideswipe, OV and PTW travelling in same directions |
| | | PTW overtaking OV while OV turning right |
| 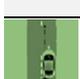 | RE/SD | PTW impacting rear of OV |
| 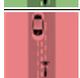 | HS/OD | head-on collision of PTW and OV |
| | | sideswipe, OV and PTW travelling in opposite directions |
| 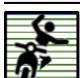 | SV‡ | PTW falling on roadway, no OV involvement |
| | | PTW running off roadway, no OV involvement |
| | | other PTW accidents with no OV or other involvement |
| 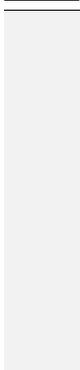 | OTHER | PTW falling on roadway in collision avoidance with OV |
| | | OV impacting rear of PTW |
| | | PTW impacting environmental object |
| | | other PTW/OV impacts |
| | | PTW impacting pedestrian or animal |
| | | PTW turning L in front of OV, OV proc in either direction perpendicular to PTW path |
| | | OV entering roadway failing to yield to PTW right of way |
| | | PTW running off roadway in collision avoidance with OV |
| | | PTW turning R in front of OV, OV proc in either direction perpendicular to PTW path |
| | | Other |

*\* OV: Other vehicle driver.*

**Table A- 2. Categories of variable primary crash contributing factor in MAIDS data. N=663** (two cases with this variable not assessed).

| PRIMARY CRASH CONTRIBUTING FACTOR | | Total (N=663) | SCP/LD (N=136) | TIP/LD (N=100) | TAP/OD (N=95) | TAP/SD (N=134) | RE/SD (N=52) | HS/OD (N=59) | SV (N=89) |
|---|---|---|---|---|---|---|---|---|---|
| OV | detection | 40.8% | 47.8% | 56.0% | 67.4% | 50.0% | 21.2% | 13.6% | 0.0% |
| | decision | 12.2% | 12.5% | 16.0% | 15.8% | 15.7% | 9.6% | 11.9% | 0.0% |
| | comprehension | 1.5% | 1.5% | 0.0% | 2.1% | 3.0% | 0.0% | 3.4% | 0.0% |
| | execution | 0.2% | 0.0% | 0.0% | 0.0% | 0.0% | 1.9% | 0.0% | 0.0% |
| PTW | detection | 9.3% | 9.6% | 2.0% | 3.2% | 11.9% | 28.8% | 8.5% | 9.0% |
| | decision | 13.1% | 11.8% | 9.0% | 3.2% | 14.2% | 11.5% | 25.4% | 21.3% |
| | comprehension | 2.4% | 3.7% | 3.0% | 2.1% | 3.0% | 3.8% | 0.0% | 0.0% |
| | execution | 4.8% | 1.5% | 0.0% | 1.1% | 0.0% | 5.8% | 13.6% | 20.2% |
| | unknown type | 2.4% | 0.0% | 0.0% | 0.0% | 0.0% | 0.0% | 1.7% | 16.9% |
| view obstruction | | 4.4% | 4.4% | 12.0% | 4.2% | 0.7% | 1.9% | 6.8% | 1.1% |
| O T H E R | adverse weather | 1.2% | 0.0% | 0.0% | 0.0% | 0.0% | 0.0% | 1.7% | 7.9% |
| | roadway maintenance defect | 0.8% | 0.0% | 0.0% | 0.0% | 0.0% | 0.0% | 0.0% | 5.6% |
| | roadway design defect | 0.6% | 1.5% | 0.0% | 1.1% | 0.0% | 0.0% | 1.7% | 0.0% |
| | roadside environ. factor, incl. animal and pedest. involv. | 0.3% | 0.0% | 0.0% | 0.0% | 0.0% | 1.9% | 0.0% | 1.1% |
| | traffic control problem, temporary traffic obstruction | 0.3% | 0.7% | 0.0% | 0.0% | 0.0% | 1.9% | 0.0% | 0.0% |
| | some manouv. of OV, not involved in the collis. | 0.3% | 0.0% | 0.0% | 0.0% | 0.0% | 1.9% | 1.7% | 0.0% |
| | OV avoiding a different collision | 0.3% | 0.0% | 0.0% | 0.0% | 0.7% | 0.0% | 1.7% | 0.0% |
| | PTW avoiding a different collision | 0.2% | 0.0% | 0.0% | 0.0% | 0.0% | 1.9% | 0.0% | 0.0% |
| | OV post-crash motions from immediate prior collision | 0.2% | 0.0% | 0.0% | 0.0% | 0.0% | 1.9% | 0.0% | 0.0% |
| | pre-existing PTW maint. related problem | 0.5% | 0.0% | 0.0% | 0.0% | 0.0% | 0.0% | 0.0% | 3.4% |
| | other | 4.5% | 5.1% | 2.0% | 0.0% | 0.7% | 5.8% | 8.5% | 13.5% |
| **TOTAL** | | 100.0% | 100.0% | 100.0% | 100.0% | 100.0% | 100.0% | 100.0% | 100.0% |

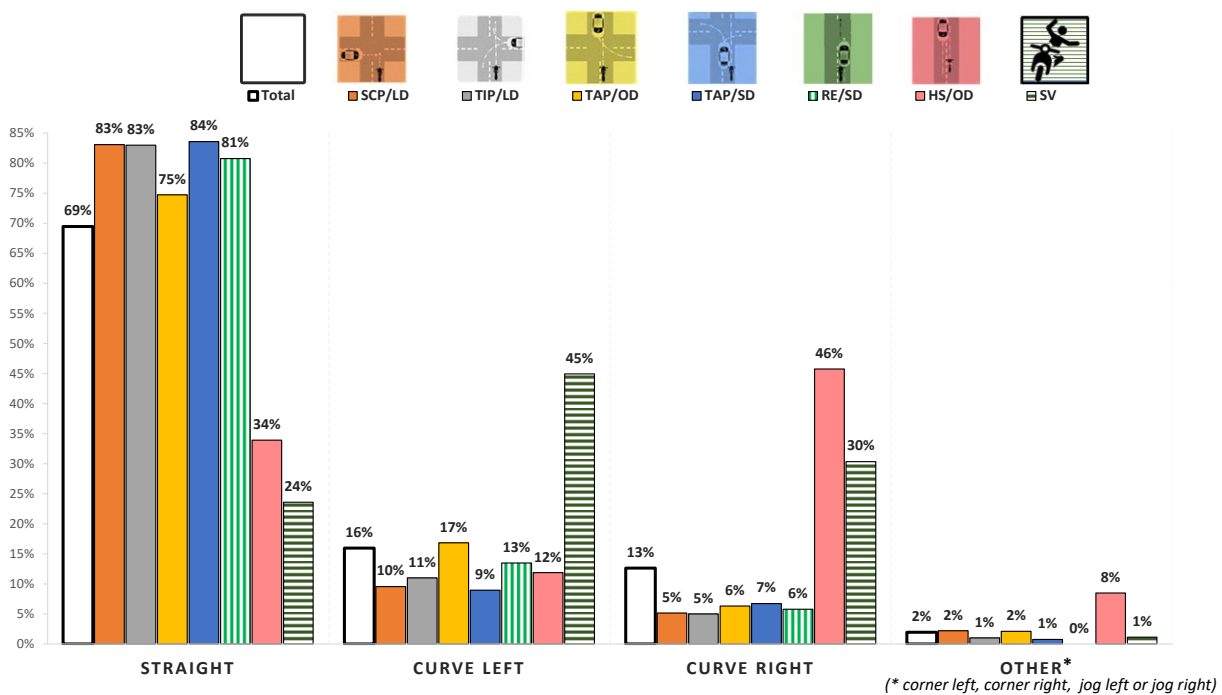

**Figure A-2. Road Alignment type per crash configuration.**

**Table A- 3. Performance of Selection of Evasive Manoeuvre when the response was not 'no evasive action'.**

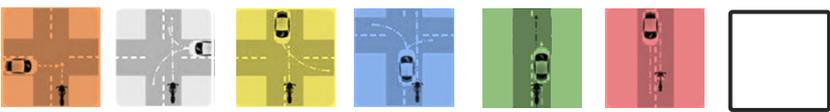

| AVOIDING CRASH MANOEUVRE | SCP/LD | TIP/LD | TAP/OD | TAP/SD | RE/SD | HS/OD | Total |
|---|---|---|---|---|---|---|---|
| **Selection** | N=72 | N=81 | N=65 | N=78 | N=41 | N=37 | N=374 |
| not properly selected | 19.4% | 16.0% | 10.8% | 16.7% | 4.9% | 21.6% | 15.2% |
| properly selected | 76.4% | 82.7% | 86.2% | 79.5% | 87.8% | 78.4% | 81.6% |
| unknown | 4.2% | 1.2% | 3.1% | 3.8% | 7.3% | 0.0% | 3.2% |
| TOTAL | 100.0% | 100.0% | 100.0% | 100.0% | 100.0% | 100.0% | 100.0% |

**Table A- 4. Performance of Execution of Braking or Swerving as Evasive Manoeuvre for cases with the proper selection.**

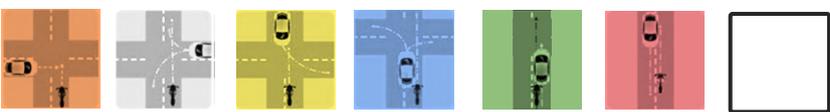

| AVOIDING CRASH MANOEUVRE | SCP/LD | TIP/LD | TAP/OD | TAP/SD | RE/SD | HS/OD | Total |
|---|---|---|---|---|---|---|---|
| **Braking** | N=41 | N=53 | N=46 | N=43 | N=26 | N=16 | N=225 |
| not properly executed | 43.9% | 39.6% | 45.7% | 41.9% | 46.2% | 43.8% | 43.1% |
| properly executed | 46.3% | 54.7% | 52.2% | 55.8% | 53.8% | 56.3% | 52.9% |
| unknown | 9.8% | 5.7% | 2.2% | 2.3% | 0.0% | 0.0% | 4.0% |
| TOTAL | 100.0% | 100.0% | 100.0% | 100.0% | 100.0% | 100.0% | 100.0% |
| **Swerving** | N=10 | N=10 | N=6 | N=16 | N=8 | N=8 | N=58 |
| not properly executed | 20.0% | 40.0% | 50.0% | 12.5% | 50.0% | *87.5% | 37.9% |
| properly executed | 80.0% | 50.0% | 50.0% | 81.3% | 37.5% | 12.5% | 56.9% |
| unknown | 0.0% | 10.0% | 0.0% | 6.3% | 12.5% | 0.0% | 5.2% |
| TOTAL | 100.0% | 100.0% | 100.0% | 100.0% | 100.0% | 100.0% | 100.0% |
| **Other** * | N=4 | N=4 | N=4 | N=3 | N=2 | N=5 | N=22 |
| not properly executed | 25.0% | 50.0% | 25.0% | 100.0% | 0.0% | 40.0% | 40.9% |
| properly executed | 75.0% | 50.0% | 75.0% | 0.0% | 100.0% | 40.0% | 54.5% |
| unknown | 0.0% | 0.0% | 0.0% | 0.0% | 0.0% | 20.0% | 4.5% |
| TOTAL | 100.0% | 100.0% | 100.0% | 100.0% | 100.0% | 100.0% | 100.0% |

*\* lane changing, accelerating, flashing headlamp high beams, drag feet, honking or unknown*

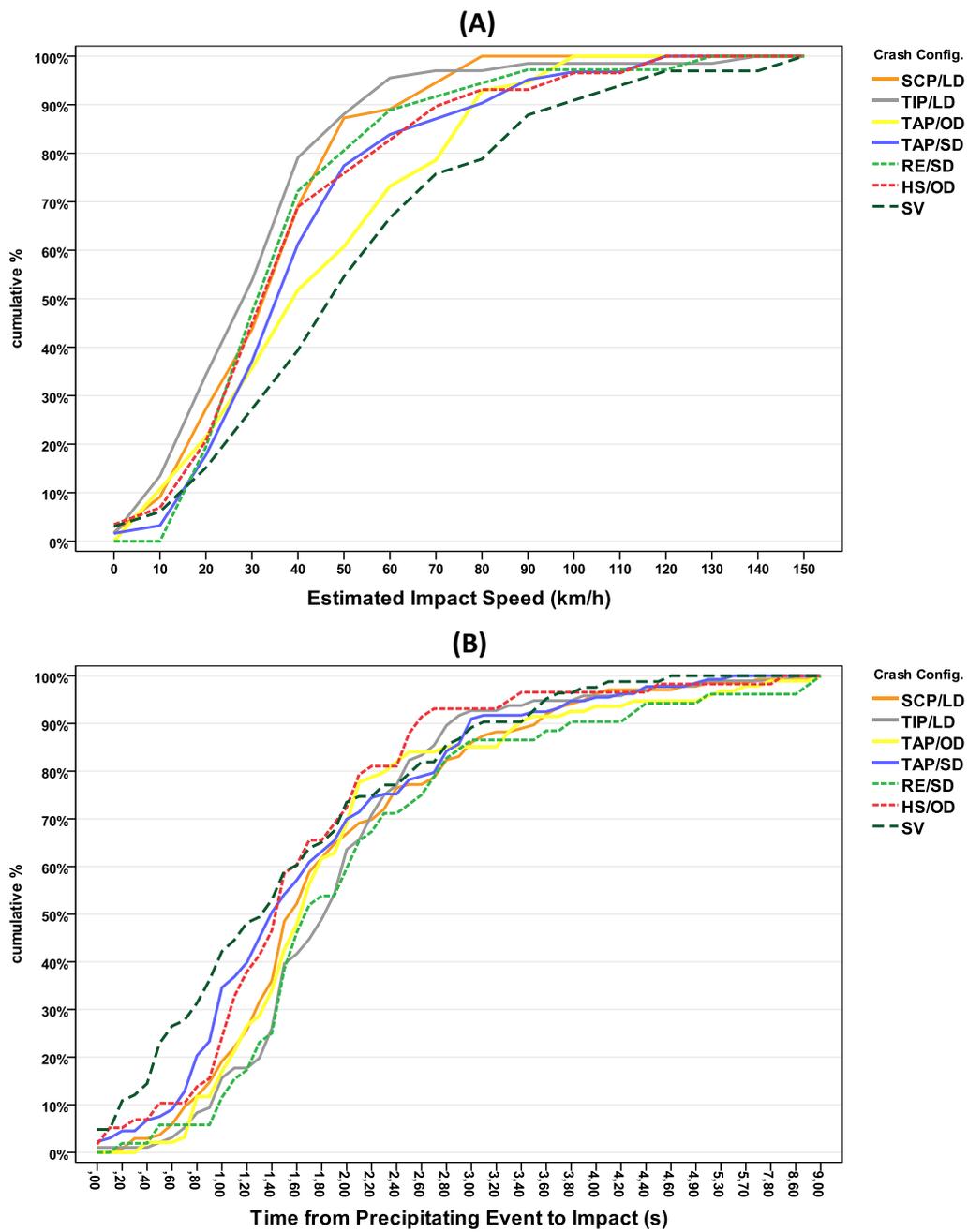

**Figure A- 3. Cumulative distribution per crash configuration type for (A) 'Impact Speed' and (B) 'TPEI'**

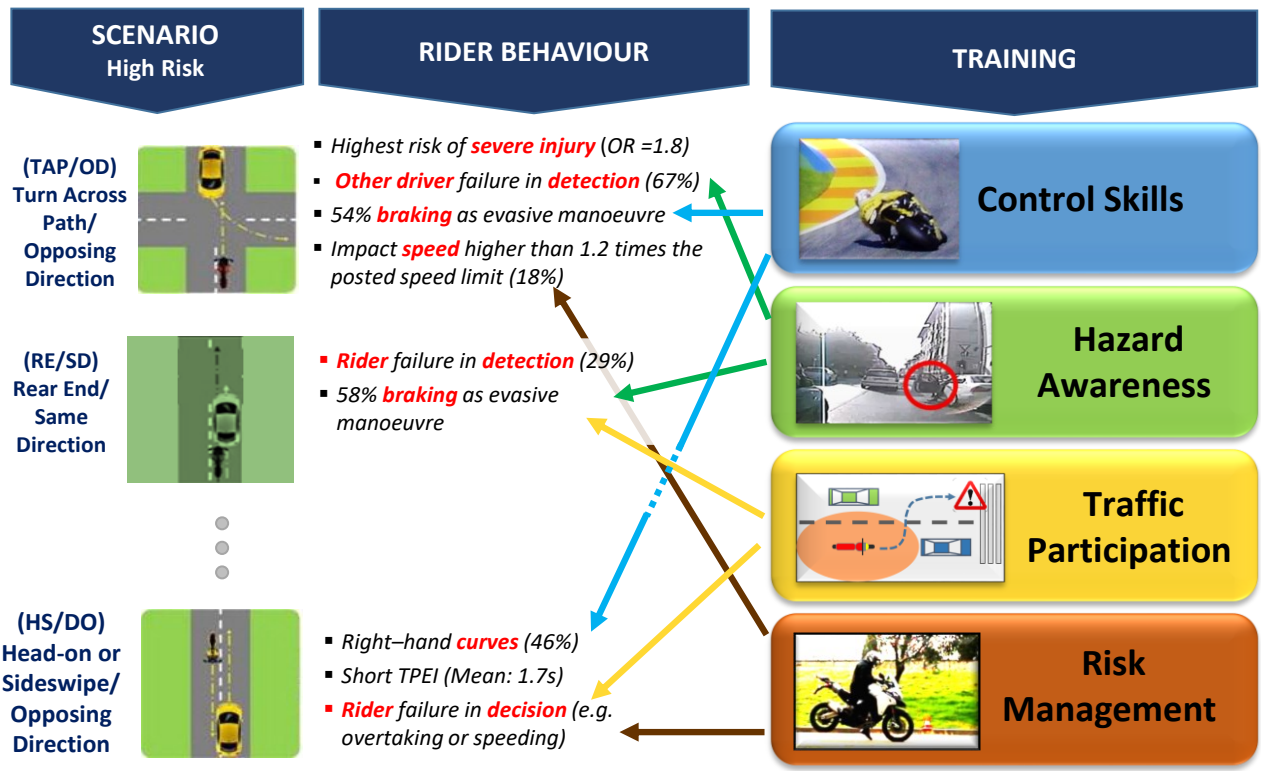

**Figure A- 4. Summary of Methodology to address training of the skills associated with high risk scenarios.**

## ADDITIONAL BIBLIOGRAPHY